\documentclass[aip,apl,reprint,amsmath,amssymb]{revtex4-1}

\usepackage{graphicx}%
\usepackage{dcolumn}%
\usepackage{bm}%
\usepackage{units}%

\begin{document}

\title{Insight into bulk niobium superconducting RF cavities performances by tunneling spectroscopy}

\author{N.R. Groll}
\affiliation{Argonne National Laboratory, Material Sciences Division, 60439, Lemont, USA.}
\author{C. Becker}
\affiliation{Illinois Institute of Technoloy, Physics Departement, 60616 Chicago, USA.}
\author{G. Ciovati}
\affiliation{Thomas Jefferson Laboratory, Accelerator Operations Division, 23606 Newport News, USA.}
\author{A. Grassellino}
\affiliation{Fermi National Accelerator Laboratory, Technical Division, 60510 Batavia, USA.}
\author{A. Romanenko}
\affiliation{Fermi National Accelerator Laboratory, Technical Division, 60510 Batavia, USA.}
\author{J.F.Zasadzinski}
\affiliation{Illinois Institute of Technoloy, Physics Departement, 60616 Chicago, USA.}
\author{T.Proslier}
\email[]{thomas.proslier@cea.fr}
\affiliation{IRFU, CEA, University Paris-Saclay, F-91191 Gif-sur-Yvette, France}

\date{\today}

\begin{abstract}
Point contact tunneling (PCT) spectroscopy measurements are reported over wide areas of cm-sized cut outs from superconducting RF cavities prepared on Niobium.  A comparison is made between a high-quality, conventionally processed (CP) cavity with a high field Q drop for acceleration field E $>$ 25 MV/m  and a nitrogen doped (N-doped) cavity that exhibits an increasing Q up to fields approaching 15 MV/m. The CP cavity displays hot spot regions at high RF fields where Q-drop occurs as well as unaffected regions (cold spots). PCT data on cold spots reveals a near ideal BCS density of states (DOS) with gap parameters, $\Delta$ as high as 1.62 meV, that are among the highest values ever reported for Nb. Hot spot regions exhibit a wide distribution of gap values down to $\Delta \sim$ 1.0 meV and DOS broadening characterized by a relatively large value of pair-breaking rate, $\Gamma$,  indicating surface regions of significantly reduced superconductivity. In addition, hot spots commonly exhibit Kondo tunneling peaks indicative of surface magnetic moments attributed to a defective oxide.  N-doped cavities reveal a more homoegeneous gap distribution centered at $\Delta \sim$ 1.5 meV and relatively small values of $\Gamma/\Delta$.  The absence of regions of significantly reduced superconductivity indicates that the N interstitials are playing an important role in preventing the formation of hydride phases and other macroscopic defects which might otherwise severely affect the local, surface superconductivity that lead to hot spot formation. The N-doped cavities also display a significantly improved surface oxide, i.e., increased thickness and tunnel barrier height, compared to CP cavities. These results help explain the improved performance of N-doped cavities and give insights into the origin of the initial increasing Q with RF amplitude.
\end{abstract}
\maketitle

\section{Introduction}
\begin{figure}
	\centering
	\includegraphics[width=0.5\textwidth]{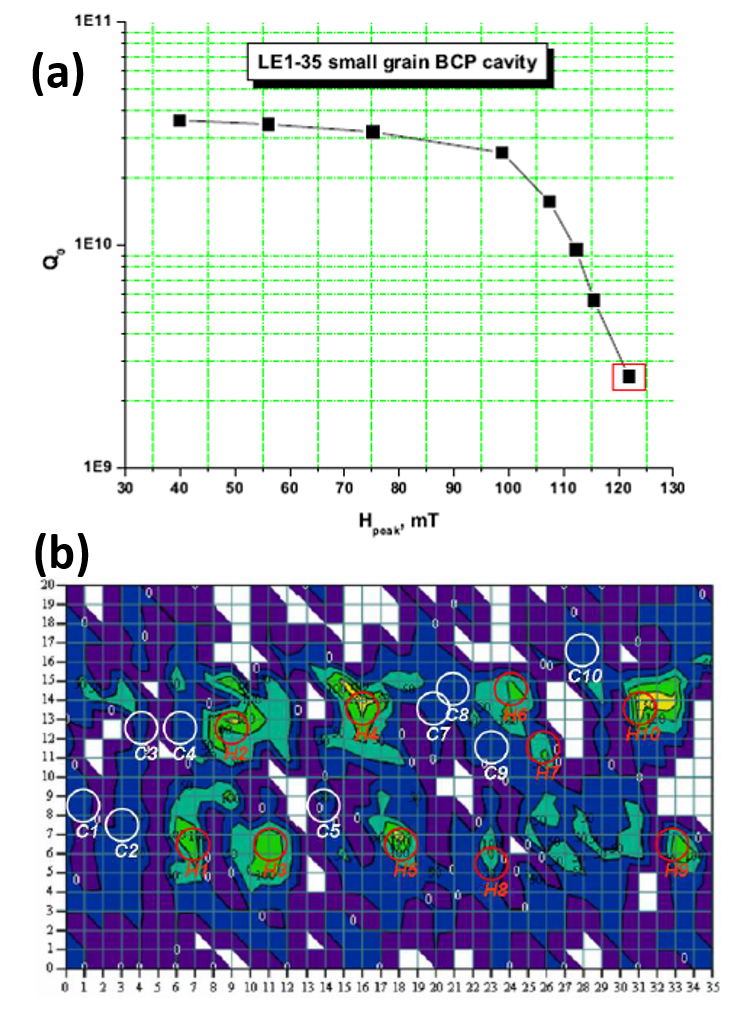}
	\caption{(Color online) (a) RF test at 1.4 K of a 1.3 GHz monocell Tesla-shaped small grain Nb cavity that have been BCPed. (b) temperature map measured at 120 mT (corresponding red square in (a)). The samples cut out from this cavity are C8 and H2 and indicated with arrows. The figures were reproduced from Ref.\cite{RomanenkoThesis}.}
	\label{fig:RF performances1}
\end{figure}

Superconducting Radio frenquency (SRF) cavities are the corner stone infrastructure of most actual and future particule accelerators. Steady technological developments over the past 50 years have enabled reproducible state of the art SRF cavity performance with quality factors, Q, of $\sim 10^{10}$ and maximum accelerating fields, E$_{MAX}$, of $\sim 35$ MV/m\cite{Padamsee}. A typical RF performance test of a buffered chemical polished (BCP) Tesla-shaped 1.3 GHz niobium (Nb) cavity\cite{RomanenkoThesis}, represented in Fig.\ref{fig:RF performances1}(a), reveals characteristic features of a nearly field-independent Q until a peak surface field, H$_{Peak}\sim$ 100 mT followed by a drastic decrease, so-called the high field Q slope (HFQS), until a maximum H$_{PeakMax}$ value of 120 mT. Such Q-slope is indicative of strong dissipative processes occuring in the high field regime. The temperature maps measured at a H$_{Peak}$ of 120 mT on the cavity walls consistently show localized regions of strong temperature increase, labeled as hot spots in green in Fig.\ref{fig:RF performances1}(b), surrounded by area of lesser dissipation named cold spots. Similar trends are reproducibly found on electropolished (EP) Nb cavities (black curve in Fig.\ref{fig:RF performances2}(a)) and were considered as a performance standard until recently.

Original works done at Fermilab and Jlab showed that, high temperature treatments (HT) under controlled N$_2$ atmosphere \cite{AnnaNDoping1,AnnaNDoping2} or high vacuum (HV) \cite{Dhakal1} on Nb cavities, cause a pronounced increase of the quality factor with increasing magnetic field intensity, so-called anti Q slope, until a quench around 70-100 mT (Fig.\ref{fig:RF performances2}). The improvement of Q $\propto 1/R_S$ implies a significant reduction of the surface impedance, R$_S$, with increasing RF field and induced screening supercurrent intensity. This unexpected result triggered intensive experimental and theoretical efforts\cite{AnnaNDoping1,AnnaNDoping2,Gurevich} aimed at understanding the microscopic origins of such phenomena as well as defined a new standard for SRF Nb cavities performance.

The RF performances strongly depend on the Nb superconducting properties within a few penetration depth ($\lambda\sim$ 40 nm) from the cavity surface exposed to RF fields. It is therefore crucial to understand how different surface treatments (chemical etching, doping, annealing...) affect the surface superconducting parameters such as the superconducting gap, $\Delta$, the critical temperature, $T_C$, and non-ideal superconducting signatures (proximity effect, inelastic scattering processes) relevant to RF performance limitations. In addition to the superconducting characteristics, previous work\cite{Q-bits1,Q-bits2,Q-bits3,TLS-1,TLS-2,TLS-3} underlined the importance of the dielectric native oxide  properties on the performances limitations of superconducting planar resonators in the GHz regime. 

Tunneling spectroscopy (TS) and in particular point contact tunneling spectrocopy (PCTS) is a versatile technique that uses the substrate native oxide as the tunnel barrier to probe the surface superconducting density of states (DOS). The contact regime hence offers a unique opportunity to study both the surface superconductivity and the dielectric layer properties of cut out or samples that mimic processes applied to real devices including SRF cavities\cite{proslier2015InstrumPCT}.

This work provides a comparative study of the surface DOS and the tunnel barrier properties (thickness and work function) measured over wide surface areas by means of PCTS between samples cut out from an optimized nitrogen-processed cavity, and a hot and cold spot in the HFQS region of a BCP cavity\cite{RomanenkoThesis,AnnaNDoping1}. The correlations between the surface structural, chemical, and electronic properties obtained using complementary surface characterization technics on Nb cut out samples and the corresponding RF performance cavity tests enable the identification of potential causes (deleterious phases, magnetic impurities) and relevant parameters for RF performance limitations and optimization. 

\section{Experimental Results}

The superconducting DOS measured by tunneling spectroscopy with a normal metal tip can be extracted from the tunneling conductance curve from the relation:
\begin{equation}
\frac{dI(V)}{dV}\propto \int^{\infty}_{-\infty}P(E)N_{S}(E,\Gamma,\Delta)\left[ -\frac{\partial f(E+eV)}{\partial (eV)} \right]dE,
\end{equation}

Where $P(E)$ is the tunneling probability, $f(E)$ is the Fermi distribution function, $N_{S}$ is the superconducting quasiparticle density of states\cite{BCS} in the sample and given by $\frac{E-i\Gamma}{\sqrt{(E-i\Gamma)^{2}-\Delta^{2}}}$, $\Gamma$ is the phenomenological quasi-particle lifetime broadening parameter\cite{Dynes}. The critical temperature, $T_C$, is infered from the temperature dependence of the conductance spectra; the superconducting gap, ascertained from the fits using equation 1, vanishes at T=$T_C$.

For electron energies E much smaller than the tunnel barrier work function $\Phi$, $P(E)$ can be approximate as constant and taken out of the integral. Its contribution cancels out when normalizing the conductance curve $\frac{dI}{dV}$ by its value at a voltage value, V $\sim$ 10$\Delta$. In this work $\Delta\sim$ 1.5 - 1.6 meV and V=20 meV. 

The ratio $\Gamma/\Delta$ is a measure of the DOS value at the Fermi level $E_F$ (V=0 meV) that determines transport phenomena and dissipative processes in SRF cavities; for instance, previous work \cite{Kharitonov,Gurevich3,Gurevich2} have shown that the saturation of the surface impedance at low temperature, so called the residual resistance or $R_{res}$, that limits the Q value at low RF field intensity follows directly from such inelastic scattering processes that lead to a non-zero DOS at $E_F$: $\Gamma/\Delta$ is independent of temperature below T$\le T_C/3$, therefore higher $\Gamma/\Delta$ values leads to higher DOS($E_F$) and dissipation that can be measured, for instance, by thermometry. 

\begin{figure*}[ht]
	\centering
	\includegraphics[width=\textwidth]{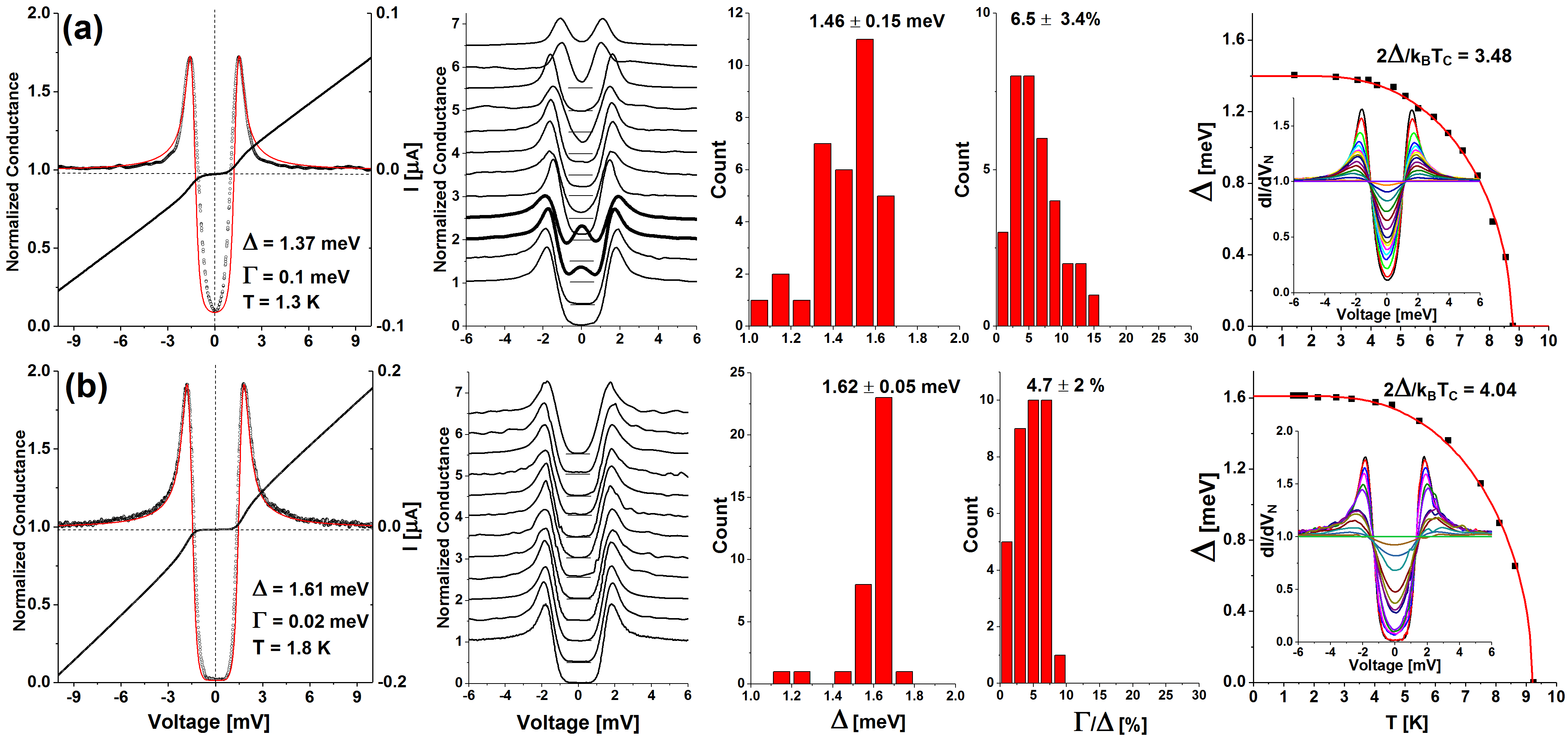}
	\caption{(Color online) Summary of the TS measurements done on the samples cut out from the hot spot H2 (a) and cold spot C8 (b) regions in Fig.\ref{fig:RF performances1}(b). from left to right: Typical normalized conductance curves dI/dV(V) with the corresponding fits, fitting parameters and current curves I(V). Series of representative conductance curves shifted for more clarity. Statistic of $\Delta$ in meV and $\Gamma/\Delta$ in $\%$ extracted from the fits on the measured conductance curves. Temperature dependence of $\Delta$ for typical tunnel junction measured and showed in the insert.}
	\label{fig:PCT HotandCold}
\end{figure*}

The $\Gamma$ parameter introduce by Dynes et al.\cite{Dynes} originally considered pair breaking processes near Tc.  However, this parameter can be treated as phenomenological to model inelastic, pair breaking processes irrespective of microscopic origin. Although more sophisticated models have been developed to quantify the effects of microscopic dissipation sources such as magnetic impurities on SRF cavities surface impedance\cite{Kharitonov}, the Dynes model captures a simple, qualitative parameter $\Gamma$, that can be used to draw general correlations between RF test performances and surface inelastic processes. Throughout this work we choose to use the parameter $\Gamma/\Delta$ in $\%$ to quantify those processes whenever pertinent.

For electron energies close to $\Phi$, $P(E)$ is not constant anymore and the conductance is sensitive to the tunnel barrier properties. Brinkman et al.\cite{Brinkman} modeled the tunnel barrier by a trapezoidal potential barrier and for large voltages, the conductance curve $G(V)=dI(V)/dV$ can be approximated by the relation:
\begin{equation}
\frac{G(V)}{G(0)}=1-\frac{A_0\Delta\phi}{16\phi_{ave}^{3/2}}eV+\frac{9A_0^2}{128\phi_{ave}}(eV)^2,
\end{equation}
Where $\Delta\phi$ and $\phi_{ave}$ are the work function difference and average between the tip and the sample. $A_0 = 4(2m)^{1/2}d/3\hbar$ where $m$ is the electron mass and d the tunnel barrier thickness. The fits of $G(V)/G(0)$ at high bias gives d in {\AA} and the sample work function $\phi$ in eV.

Repeated measurements of different tunnel junctions between 1.3 K and 1.8 K over areas of tens of microns provide a representative statistical distribution of the superconducting parameters $\Delta$, $\Gamma$, and the tunnel barrier parameters, d and $\Phi$ for each samples. In addition, for the hot spot and the nitrogen-processed samples we were able to map the spatial variations of the superconducting parameters over area with lateral size of $\sim$ 100-300 $\mu$m. 

\subsection{Hot and cold spot}

The samples studied were cut out from regions C8 (cold spot) and H2 (hot spot) indicated on the temperature map in Fig.\ref{fig:RF performances1}(b). The cavity is a medium grain Nb 1.3 GHz Tesla-shaped resonnator that has been chemicaly etched using BCP recipe\cite{RomanenkoThesis}.

\begin{figure}
	\centering
	\includegraphics[width=0.5\textwidth]{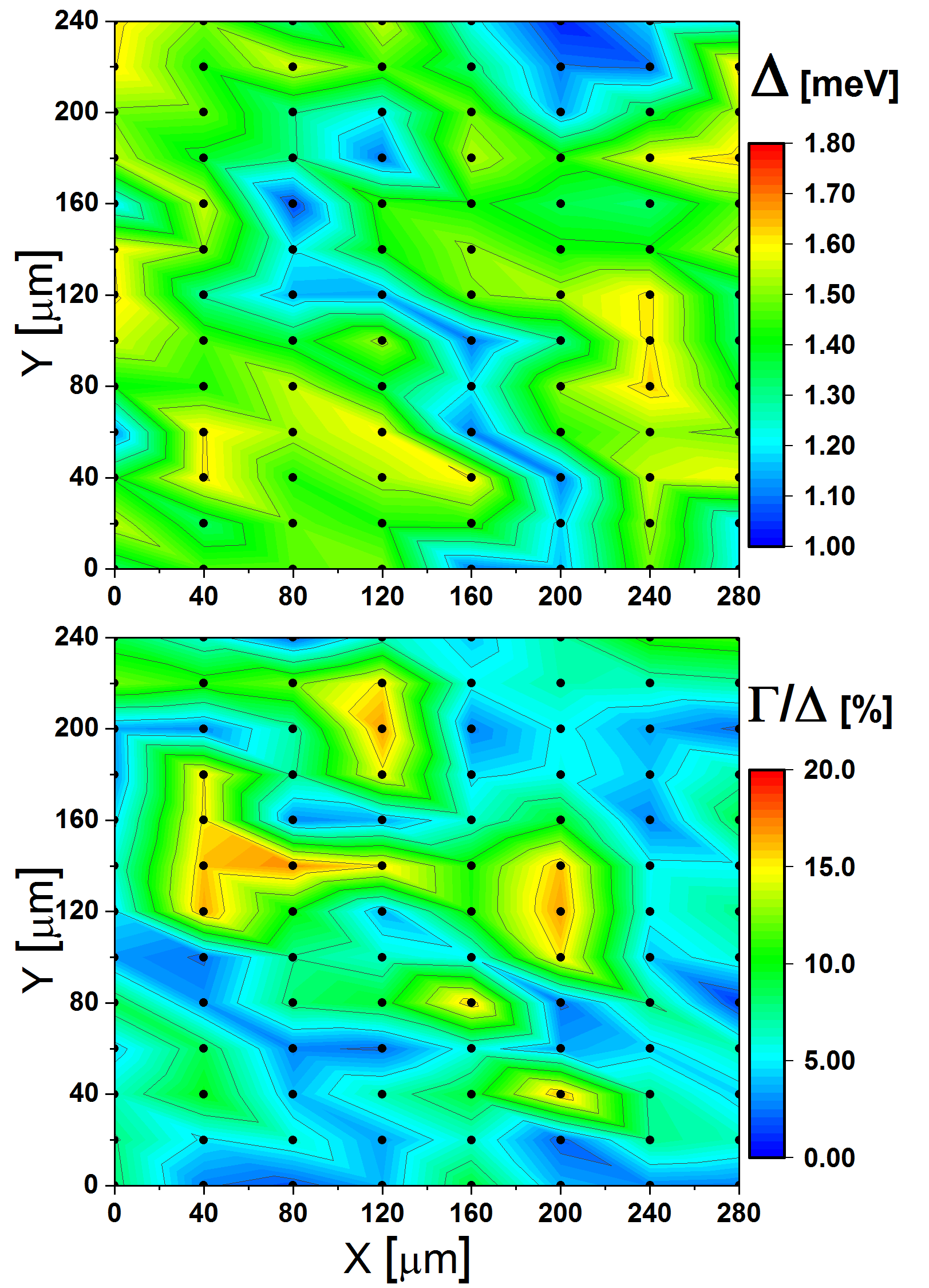}
	\caption{(Color online) Maps of the superconducting gap $\Delta$ (top) and $\Gamma/\Delta$ (bottom) of the hot spot sample extracted from the fits using equation 1. The black points represent the location of the tunnel junctions}
	\label{fig:MapHotSpot}
\end{figure}

The PCTS data measured on H2 and summarized in Fig.\ref{fig:PCT HotandCold}(a) exhibit a broad distribution of $\Delta$ values peaked at the accepted bulk Nb $\Delta_{Nb}$, between 1.5 and 1.61 meV depending on the crystalline orientation\cite{MacVicar}. A significant number of junctions however have superconducting gap values much smaller than $\Delta_{Nb}$, as low as 1 meV, and $\Gamma/\Delta$ values up to 15$\%$ indicative of a inhomogeneous sample with areas of weakened superconductivity and likely high RF dissipation. The cold spot sample (Fig.\ref{fig:PCT HotandCold}(b)), on the contrary, exhibits a very narrow gap and $\Gamma/\Delta$ distribution around 1.62 meV and 4.7$\%$ respectively, characteristic of a homogeneous, near-ideal BCS superconducting region with much weaker dissipation. These tunneling results are consistent with the temperature increase map measured at high RF field in Fig.\ref{fig:RF performances1}(b). The spatial variations of the superconducting gap $\Delta$ and $\Gamma/\Delta$ over an aera of 240 x 280 $\mu$m are represented in Fig.\ref{fig:MapHotSpot}.

Another peculiar spectral feature measured on the hot spot sample is the occurence of zero bias peaks (ZBP) around the Fermi level (bold conductance curves in Fig.\ref{fig:PCT HotandCold}(a)). These peaks, absent in the cold spot sample, are the signature of a tunneling Kondo effect\cite{Kondo} that indicates the presence of localized magnetic moments in or at the vicinity of the native oxide tunnel barrier. Such features have been previously measured on highly dissipative Nb samples\cite{Proslier_APL_2008,Proslier_2011}. The $\Gamma/\Delta$ values extracted from the quasiparticle peaks fits of these junctions are between 10 to 15$\%$ and account for 60$\%$ of the statistical weight at high $\Gamma/\Delta$ values. This result suggests that magnetic impurities contribute to dissipative processes in hot spots measured in the high field Q slope region. 

\begin{figure}
	\centering
	\includegraphics[width=0.5\textwidth]{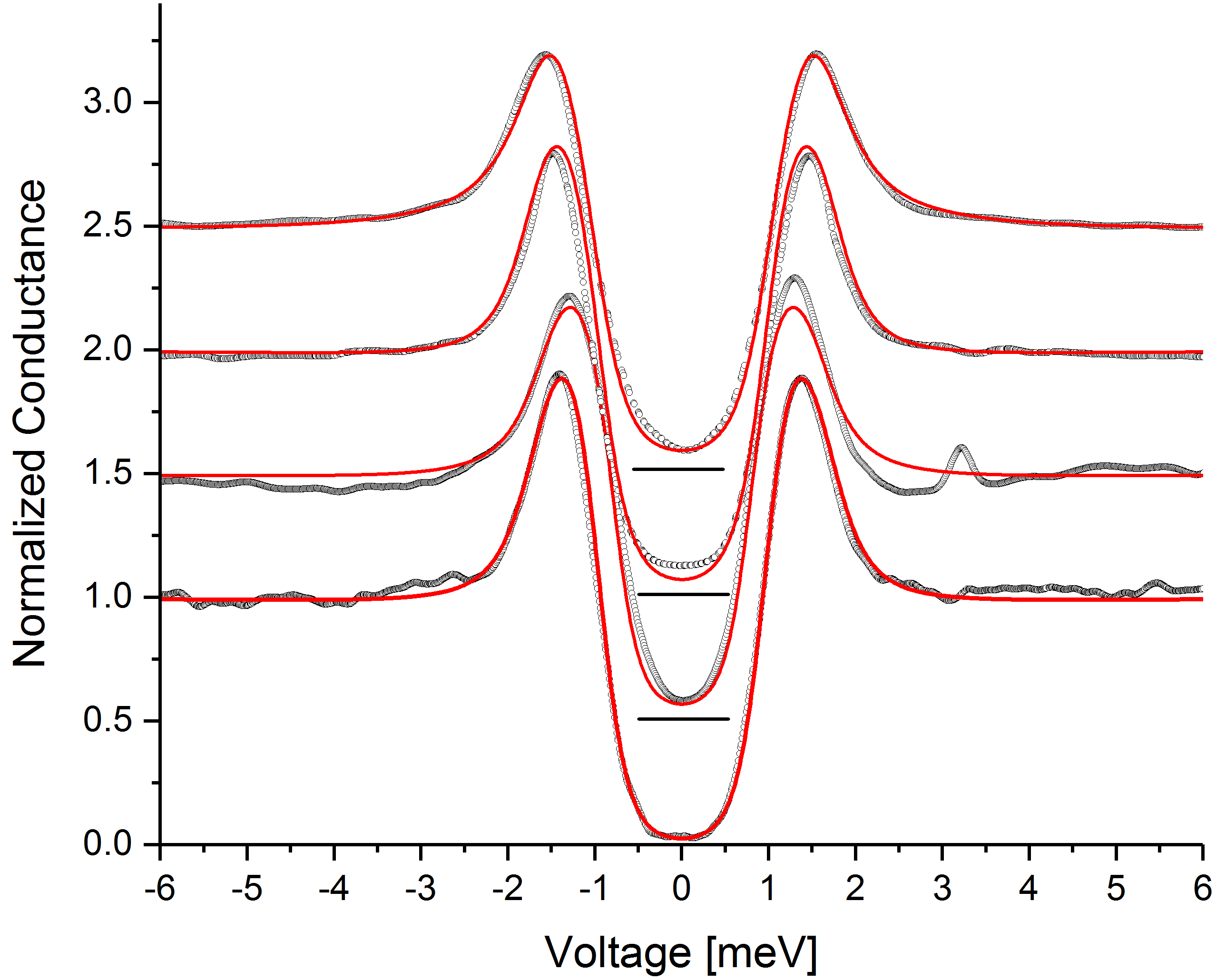}
	\caption{(Color online) Selected tunneling conductance curves and the corresponding fits in red using the McMillan theory from samples H2 (top 2 curves) and C8 shifted for more clarity. The parameters: $\Delta_N$, $\Delta_S$, $\Gamma_N$, $\Gamma_S$ and $\Gamma$ in meV extracted from the fits are from top to bottom: 0, 1.65, 4, 1, 0.08 $\setminus$ 0, 1.65, 2.7, 0.2, 0.06 $\setminus$ 0, 1.62, 2.6, 0.7, 0.05 $\setminus$ 0, 1.65, 2.8, 0.3, 0.04.}
	\label{fig:PCT McMillan}
\end{figure}

If the signature of magnetic impurities in the tunneling curves is associated with higher $\Gamma/\Delta$, the presence of tunnel junctions with small gap and small $\Gamma/\Delta$ values indicate an additional microscopic mechanism affecting the surface impedance; in the two fluids model\cite{Bardeen}, $R_S(T) = R_{BCS}(T)+R_{res}$ where $R_{BCS}$ is the BCS\cite{BCS} part of the surface impedance $\propto e^{-\Delta/(k_BT)}$ that arises from thermally excited quasiparticles. Hence small superconducting gap $\Delta$ values measured in the hot spot samples will cause an increase of $R_S$. Typical conductance spectra with gap values between 1.1 and 1.3 meV using equation 1 are represented in Fig.\ref{fig:PCT McMillan}. 

Several mechanisms can be responsible for reducing the gap below the bulk Nb value. The presence of nanometric size clusters of niobium hydrides revealed by transmission electron microscopy (TEM)\cite{TEMHydrides} in the first micron of cut out hot spots regions in the HFQS regime indicate that a proximity effect between the metallic, non-superconducting NbH$_x$\cite{Tc} and the surrounding bulk niobium might be the cause for depressed superconductivity.

In order to verify this scenario,  the McMillan model\cite{McMillan} was used to fit the small gap value conductance spectra. This model calculates the proximity effect between a superconducting film S of thickness $d_S$, order parameter $\Delta_S$, and DOS in the normal state $N_S(0)$ separated by a potential barrier from and a normal metal N of thickness $d_N$, order parameter $\Delta_N$, and DOS in the normal metal $N_N(0)$. The superconducting parameters are assumed to be uniform accross the SN sandwitch and the quasi-particle scattering at the interface diffusive. The tunneling rates between the normal and superconducting films are given by $\Gamma_N$ and $\Gamma_S$ and the parameter $\Gamma$ was added to account for inelastic scattering in the normal metal N. 

For all the junctions we assumed $\Delta_N$=0. The fits, shown in Fig.\ref{fig:PCT McMillan}, are in very good agreement with the data, better in fact than with the Dynes model. The best fits were obtained for ratios $\frac{\Gamma_N}{\Gamma_S}=\frac{d_S\times N_S(0)}{d_N\times N_N(0)}\gg$1, $\Gamma_S\ll\Delta_S$ and $\Gamma\sim$ 0.05 meV. Assuming that $N_S(0)\sim N_N(0)$ we have that $d_N \ll d_S$ indicative of a thin normal layer or cluster on top of bulk Nb. We can estimate $d_N$ given the relation $d_N=(\hbar\times v_F\times \sigma)/(2\times B\times \Gamma_N)$ where $v_F$ is the fermi velocity in the normal metal that we took as $\sim v_F$ in niobium $=1.4\times10^{6}$ m/s from ref. \cite{Weber}, $\sigma$ is the barrier penetration probability assumed to be $\le$ 0.1 and B is a function of the ratio of the mean free path, $l$, to the film thickness $d_N$. Following McMillan, B $=2$ for $l\sim d_N$ and we obtain that $d_N\le$ 14 nm. The normal metal layer thickness estimate is consistent with the nanometric Nb$H_x$ cluster size measured by TEM\cite{TEMHydrides}.

We attempted fitting the same spectra with the Arnold model\cite{Arnold} that assumes a perfect contact ($\sigma\sim$1) between the normal and superconducting metals but the agreement with the data was not as good as with the McMillan model. The low interface transparency as well as the diffusive nature of quasiparticle tunneling assumed in the McMillan model could be explained by the presence of a non-negligeable interstitial concentration, for instance hydrogen, at the intergace between the normal metal cluster and the surrounding Nb. It could also be an indication that the interface between Nb and NbH has a large lattice mismatch and consequent large diffusive scattering.

The temperature dependence of tunneling junction with bulk Nb gap values (Fig.\ref{fig:PCT HotandCold} (b)) reveals $T_C$ between 9.25 and 9.3 K, typical of bulk niobium. The junction measured on the hot spot sample, with a $\Delta=$1.39 meV shows a slightly depressed $T_C$ value of 8.8$\pm$0.2 K consistent with the calculated $T_C$ of 8.4K using the McMillan formula:

$ln(T_C^B/T_C)=\frac{\Gamma_S}{\Gamma_S+\Gamma_N}(\Psi(\frac{\Gamma_N+\Gamma_S}{\pi.k_B.T_C}+0.5)-\Psi(0.5))$  

where $T_C^B$ is the bulk Nb $T_C$ of 9.3 K, $\Psi$ is the digamma function, $\Gamma_S$ and $\Gamma_N$ = 0.2 and 2.7 meV as infered from the fit of the corresponding conductance curve (insert Fig.\ref{fig:PCT HotandCold} (b), right).

X-ray diffraction (XRD) measurements were performed on sample H2 down to cryogenic temperatures in order to substantiate the presence of Nb hydride phases. The experiment was done at the Advance Photon Source (APS), beamline 33BM, with a beam energy of 10 keV. The sample was kept at 90 K for 3 hours prior to decreasing the cryogenic stage temperature down to 50K at which point the beam shutter was open and diffraction data collected. The diffraction data and pole figures measured show no sign of crystalline phases other than pure Nb. It is however possible that the sample exposure to intense photon flux induce the hydride phases dissociation too quickly to be measured. The elusive and fragile nature of niobium hydrides have also been observed in the TEM cross section experiments mentioned previously\cite{TEMHydrides2}.

In summary, the BCP Nb cavity shows inhomogeneous superconductiving properties with regions of high dissipation. We have identified two microscopic phenomena possibly responsible for the dissipation measured at high RF field intensity: magnetic impurities and the proximity effect. The localized magnetic moments revealed by the kondo signature at the Fermi level can originate from two level systems (TLS) located next or within the native oxide. The good agreement between the proximity effect McMillan model and the low gap value conductance spectra unveil the presence of deleterious clusters with a typical size $<$ 14 nm present within the penetration depth, $\lambda\sim$40 nm of Niobium in hot spot regions. At present niobium hydrides are likely candidates, but other impurities phases such as amorphous carbon or nanometric niobium carbide that have been measured in highly dissipative Nb samples\cite{NbCarbide1,NbCarbide2} could also be responsible for the HFQS.

\subsection{N doped sample}

 \begin{figure}
	\centering
	\includegraphics[width=0.5\textwidth]{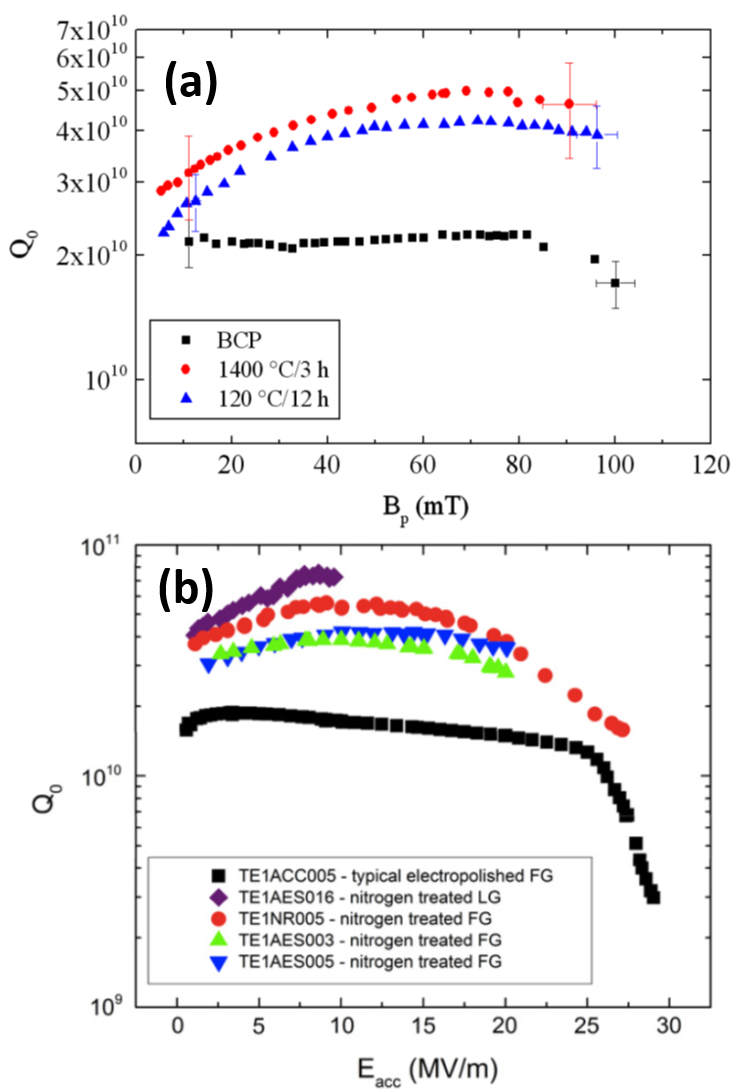}
	\caption{(Color online) (a) Q$_0$ vs $B_{Peak}$ measured after etching by BCP followed by HT at 1400$^{\circ}$C and a low temperature baking at 120$^{\circ}$C. (b) Q$_0$ vs $E_{Acc}$ measured after EP and followed by Nitrogen treatment. The sample was cut out from the cavity corresponding to the round dots. Figures (a) and (b) were reproduced from Ref.\cite{AnnaNDoping1} and \cite{Dhakal1}.}
	\label{fig:RF performances2}
\end{figure}

\begin{figure*}
	\centering
	\includegraphics[width=\textwidth]{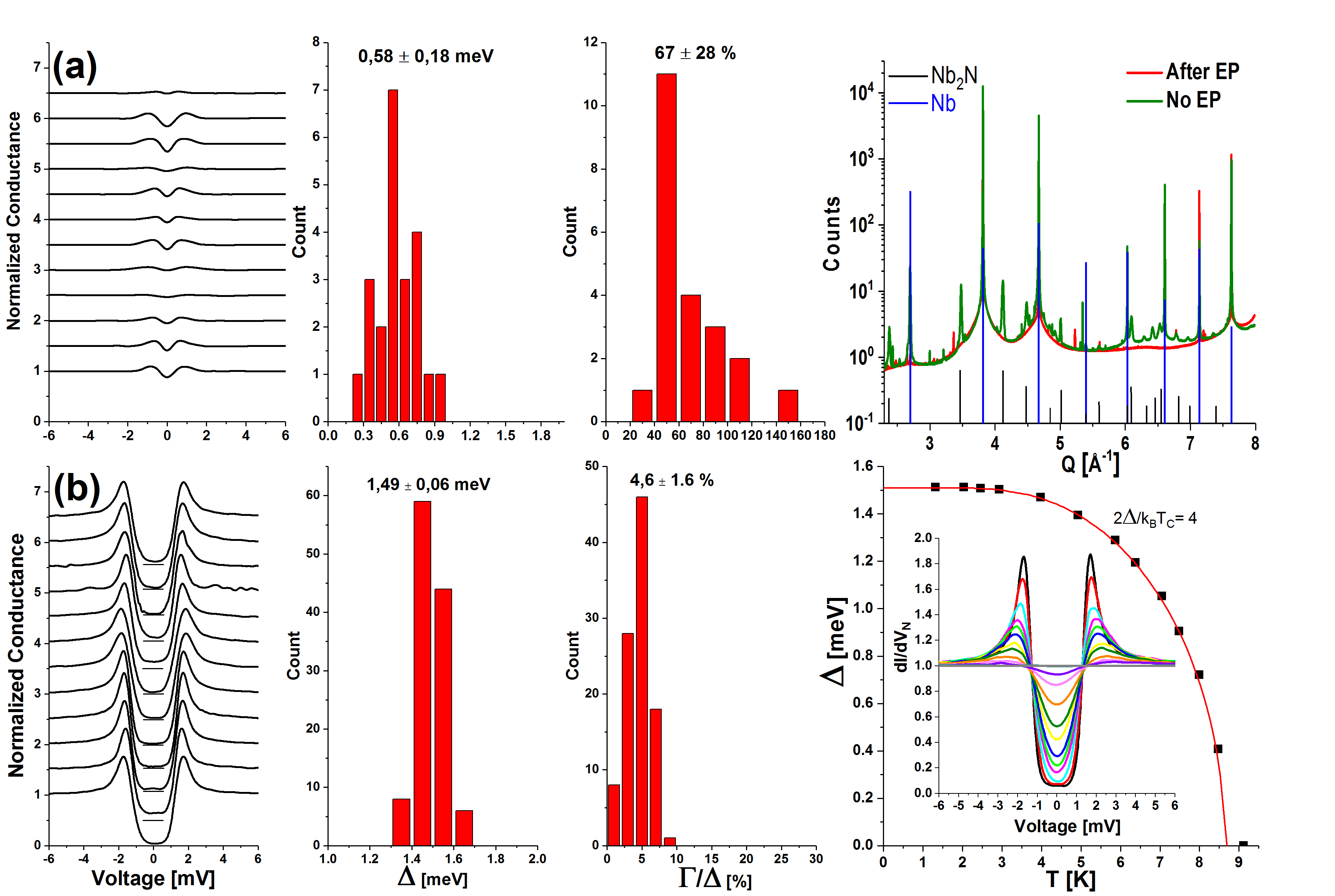}
	\caption{(Color online) Summary of the TS measurements done on the Nb samples cut out from a nitrogen processed Nb cavity before EP (a) after 10 $\mu$m EP with the corresponding RF test shown in Fig.\ref{fig:RF performances2}(b) in red round dots. from left to right: Series of representative conductance curves shifted for more clarity. Statistic of $\Delta$ in meV and $\Gamma/\Delta$ in $\%$ extracted from the fits of the measured conductance curves. (Right top) X-ray diffraction spectra of the same two samples along with the Nb$_2$N and NbN diffraction patterns.(Right bottom) Temperature dependence of $\Delta$ for typical tunnel junction measured and showed in the insert.}
	\label{fig:PCT NDopedNb}
\end{figure*}

We now focus on the newly developed HT processes found to increase the SRF cavity Q value by a factor of 3 to 5 with increasing peak surface RF magnetic field intensity from 0 to $\sim$ 70-100 mT. The medium grain 1.3 GHz niobium cavity was annealed in ultra high vacuum (UHV) for 3hr at 800$^{\circ}$C followed by 2 min at the same temperature in 25 mTorr of N$_2$. The cavity has been subsequently electropolished (EP) to remove 5$\mu$m of material prior to the RF test shown in Fig.\ref{fig:RF performances2}(b). A sample was cut out from the cavity labeled TE1NR005 represented in round dots in Fig.\ref{fig:RF performances2}(b) and measured by PCT. Another sample followed the same treatment without the electropolishing step.

The results obtained by tunneling spectroscopy are summarized in Fig.\ref{fig:PCT NDopedNb}. Without electropolishing, the superconducting gap values measured are very low $\sim$ 0.58 $\pm$ 0.2 meV $\ll\Delta_{Nb}$ and indicative of a strongly deteriorated superconducting surface layer most certainly caused by the nitridation process. X-ray diffraction spectra measured on the same sample and represented in Fig.\ref{fig:PCT NDopedNb}(a) right confirm the presence of polycrystalline Nb$_2$N phase with characteristic grain size and strain of 38$\pm$10 nm and 10$\pm$3$\%$ deduced from the peak positions and widths using the method developped by Balzar\cite{Balzar}. Nb$_2$N is non-superconducting\cite{Tc} and the small gap values measured could also originate from a proximity effect with the Nb underneath. Following the same procedure described for the hot spot sample, the McMillan model fits and parameters (not shown) give a normal film thickness on the order of 20-35 nm consistent with the Nb$_2$N grain size from the XRD analysis. 

\begin{figure}
	\centering
	\includegraphics[width=0.5\textwidth]{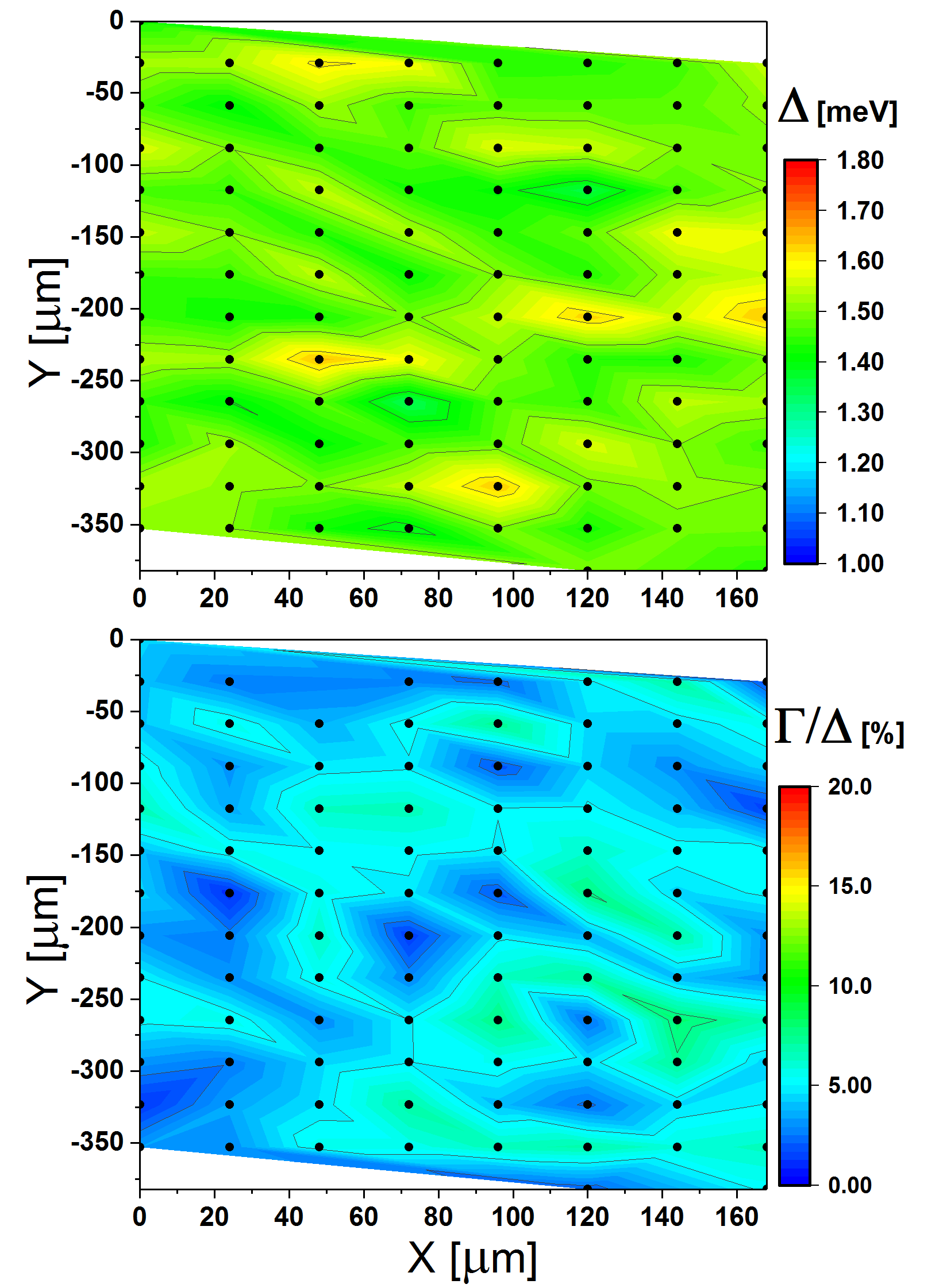}
	\caption{(Color online) Maps of the superconducting Gap, $\Delta$ (top) and $\Gamma/\Delta$ (bottom) of the N Doped sample extracted from the fits using equation 1. Each black point represent a measure of the tunneling spectra.}
	\label{fig:MapNDoped}
\end{figure}

Upon etching 5 $\mu$m of material, the tunneling junctions measured on the cut out sample exhibit high quality superconducting DOS with peaked gap and $\Gamma/\Delta$ values around 1.49 meV and 4.6$\%$ using equation 1. The maps shown in Fig.\ref{fig:MapNDoped} reveal homogeneous superconducing gap $\Delta$ and $\Gamma/\Delta$ values over a large area of 350$\times$168 $\mu$m. The averaged gap and a $T_C$ values of 8.7 K measured for a typical junction (Fig.\ref{fig:PCT NDopedNb}b) right) are slightly lower than the bulk parameters $\Delta_{Nb}$ and $T_{CNb}$. The corresponding ratio $2\Delta/k_B.T_C=4$ is in very good agreement with the ratio value of 4 extracted from the temperature dependence of the surface impedance, $R_S(T)$, measured on cavities treated similarly\cite{Martinello}. 

Surface composition analysis by secondary ion mass spectroscopy (SIMS), XRD and TEM performed on samples cut out from nitrogen treated and EP Nb cavity show the presence of interstitial N atoms with a concentration of 100-300 ppm in the first microns with no detectable Nb$_2$N phase\cite{TEMNDoped}. The slightly depressed homogeneous superconducting properties found are therefore likely due to a reduction of the niobium normal DOS($E_F$) caused by the presence of diluted interstitial nitrogen impurities, as it has been found for diluted interstitial oxygen in niobium\cite{DeSorbo}, rather than a proximity effect originating from normal metal regions near the surface.

Niobium cavities subject to a 1400$^{\circ}$C annealing in UHV for 1 hr show a very similar increase of the quality factor with increasing accelerating RF field as shown in Fig.\ref{fig:RF performances2}(a). The PCT tunneling measurement done on Nb coupons treated the same way exhibits consistently near ideal BCS DOS\cite{Dhakal1}. SIMS, X-ray photoemission spectroscopy (XPS) and XRD carried out on the same sample reveal the presence of Ti impurities, presumably originating from the NbTi cavity flanges, with concentration of a fraction of a $\%$ within the first micron into the Nb and no deleterious impurity phases or precipitates.

In summary, both HT treatments done under different conditions seems to have similar consequences; the introduction of foreign dopant atoms (N or Ti) within a few penetration depths into the niobium provides a technological passway toward an homogeneous high quality DOS with low $\Gamma/\Delta$ values. The corresponding RF tests show consistently an increase of the Q with increasing RF magnetic field intensity until a quench occuring in general at lower $H_Peak$ values than the baseline. The correlation between the tunneling spectroscopy results and the RF tests simulated recent theoretical work\cite{Gurevich} based on non-equilibrium excitation of quasi-particles driven by the high RF field and screening currents intensities. This model predicts that smaller $\Gamma/\Delta$ values and sharper quasiparticle peaks, lead to a more pronounced increase of the Q with RF field. The so-called anti-Q slope effect is therefore a signature of an homogeneous ideal BCS superconductor. For CP niobium samples the larger $\Gamma/\Delta$ values lead to an increased surface dissipation that blurs out this effect and induces a quasi field independent Q.

\subsection{Tunnel barriers}

\begin{figure*}[ht]
	\centering
	\includegraphics[width=\textwidth]{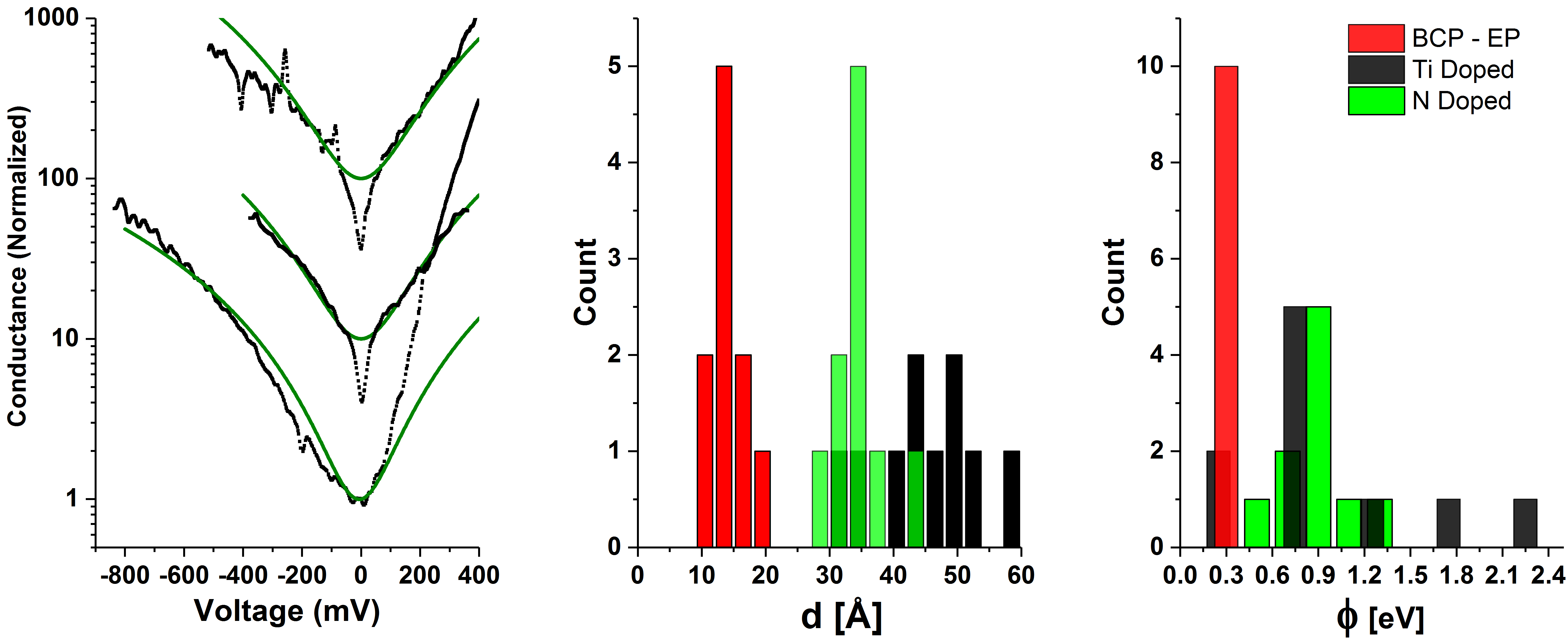}
	\caption{(Color online) Summary of the tunnel junction characteristics measured by TS. From left to right: selected high bias conductance curves with the associated fits shifted for more clarity. Statistic of tunnel barrier thickness, d, in {\AA} and work function $\phi$ in eV extracted from the fits.}
	\label{fig:PCT Barrier}
\end{figure*}

In addition to the spectroscopic signatures of the DOS at low energies $\sim\pm$ 10$\times\Delta$ around the Fermi level, we investigated the high bias region of the conductance spectra. Approaching voltages $\ge 100$ meV, the tunneling current starts reflecting the properties of the tunneling barrier itself. Typical high bias junctions measured on CP and N or Ti doped Nb samples are represented in Fig. \ref{fig:PCT Barrier}(a) along with their respective fits from equation 2. The Brinkman model has 3 parameters: The barrier thickness, d, the sample and the tip work functions, $\phi$ and $\phi_{tip}$. Depending on the metal used for the tip, Gold or Platinum-Iridium, we find a $\phi_{tip}$ of 5 and 6.1 eV in agreement with the reported values: 5.1 eV for Au and 6 eV for Pt-Ir. The statistic of d and $\Phi$ values extracted from the fits over many junctions measured on CP Nb (BCP or EP) and N or Ti doped samples and cut outs are shown in Fig.\ref{fig:PCT Barrier}. The CP Nb samples reproducibly exhibits reduced tunnel barrier thickness 15$\pm$5 {\AA} and work function 0.3$\pm$0.1 eV compared to the Ti or N doped samples with values peaked around 35$\pm$5 {\AA}, 50$\pm$10 {\AA} and 0.9$\pm$0.3 eV, 1.2$\pm$1 eV respectively. These results shows that the doping process not only affect the superconducting DOS but the native oxide tunnel barrier electronic properties as well. 

The N and Ti doped samples were measured with Au and PtIr tips that have very different hardness. The corresponding tunnel barrier fitting parameters obtained were the same for both metals which suggest that the tunnel barrier properties measured (thickness and work function) are not affected by the contact mode. 

Surface analysis using XPS, SIMS, XRD on CP Nb samples\cite{Delheusy,Halbritter} shows that the native oxide is $\sim$5 nm thick and mostly composed of amorphous Nb$_2$0$_5$ with suboxides NbO$_x$ $\sim$ 1-2 nm thick at the interface with the metallic Nb. Bulk crystalline Nb$_2$0$_5$ is an insulator with a work function of 5.1 meV, whereas the suboxides are essentially metals or semi-metals. The measured oxide tunnel barrier $\Phi\ll$ 5.1 eV indicates the presence of defects within the oxide layer. 

Foreign N or Ti atoms introduced by the HT treatments and present in the niobium prior to the oxidation via air exposure or high pressure rinsing (HPR) are consistently found in the native oxide layer by XPS and SIMS\cite{Dhakal1,TEMNDoped}. The modification of oxide electronic properties by dopant have been widely studied and used in numerous applications from light harvesting materials to catalysis. In particular, N and Ti doping \cite{OxideDoped1,OxideDoped2,OxideDoped3} have been found to affect the optical band gap, photoresponse and crystal structure of various bulk crystalline oxides including Nb$_2$0$_5$\cite{NbOxideDoped}. The increase of the tunnel barrier work function and thickness reproducibly found in the N and Ti doped Nb compared to the CP samples may originate from the neutralization of chemical or structural defects by Ti and N impurities during the oxide growth. The rare occurence of tunneling spectra with Kondo signature, presumably originating from specific defects in native amorphous oxides, in N or Ti doped samples support this scenario.

\section{Discussion}

The tunneling spectroscopy results reveal that the presence of N and Ti dopant simultaneously affect the tunnel barrier properties, the homegeneity and the quality of the superconducting DOS suggesting that a common mechanism could be responsible for all the mentioned effects.  

Upon cooling down Nb cavities, the hydrogen present in niobium and injected by the chemical etching processes (EP or BCP) can precipitate as NbH$_x$ $\alpha$, $\epsilon$ or $\beta$ phases. This phenomenum depends on the cooling rate of the cavity and is more pronounced as the cavity stays longer between 130-90 K; keeping the cavity at 100K for hours and cooling down to 2K, induce a strong degradation of the Q at low field and a pronounced Q slope above few MV/m followed by an early quench. This well known effect, so called 100K-effect\cite{Ciovati1}, is thought to be a direct consequence of hydride precipitation. The high field Q slope seen for EP or BCP cavities however is always present even for fast cooling rates $>$ 4K/min up to 10K/min which could indicate that some amount of interstitial hydrogen always precipitate as hydrides no matter how fast the cavity cooling rate is for a Nb surface layer heavily loaded with hydrogen after the chemical etching process\cite{Padamsee}. 

As a comparison, in the PCTS apparatus the samples are precooled with liquid nitrogen down to 80K with a cool down rate of 1K/min. The nitrogen gas and liquid are then evacuated for 2hrs prior to introducing the liquid Helium. The subsequent cool dow rate from 80 K down to 4.2 K is 0.3 K/min. The experiment cool down rate is therefore smaller than for the Nb cavities, promoting niobium hydride formations in samples and cut outs wherever possible. 

A low temperature baking (LTB) at 120$^{\circ}$C in UHV for 48hr on EPed or BCPed cavities, have been found to mitigate the high field Q slope and increase the quench field\cite{Visentin1,Visentin2}. The thermal maps measured at accelerating fields within the HFQS region consistently show the absence of hot spots\cite{Ciovati2}. Extensive X-ray, SIMS and TEM analysis revealed a net enrichment of subsurface interstitial oxygen by 70$\%$ within the first 10 nm during the LTB\cite{Delheusy,Ciovati1,Ciovati3} and a reduced Nb hydrides formation\cite{TEMHydrides}. Previous work from the 80's\cite{NbH1,NbH2,NbH3} showed that O or N dopants in high purity Nb samples act as traps for hydrogen atoms with 1 Hydrogen per O or N atom and a H-O(N) pair bound energy of 0.12 eV. The hydrogen trapping by N or O interstitials have been found to effectively reduce the formation of Nb hydrides\cite{NbH4}.

PCTS experiments carried out on EP and BCP Nb samples with LTB (not shown) reveal both larger mean gap values between 1.5-1.65 meV and sharper gap distribution ($\pm$ 0.07 to 0.1 meV) than the hot spot BCP Nb sample without LTB. The reduced occurence of superconducting gap values $<$ 1.4 meV on LTB sample indicate homogeneous superconducting properties and is consistent with a scenario where an increase of interstitial oxygen concentration mitigate the formation of deleterious phases (Nb hydrides) within the first penetration depth $\lambda$ of niobium.

The presence of normal surface clusters will promote flux entry by reducing the breakdown field, $H_B$ at which the Meissner effect disappear. In the proximity effect model\cite{Deutscher,Fauchere}, $H_B \approx \frac{\phi_0}{6 \lambda_N d}$ where $\phi_0$ is the flux quantum, $d$ and $\lambda_N$ are the normal metal thickness and penetration depth given by $\sqrt{m c^2/(4\pi n e^2)}$ where n is the normal metal electron density. The HFQS onset $H_B\sim$ 100 mT and n for NbH$_2$ $\sim 7.10^{28}$ m$^{-3}$ from ref.\cite{Gupta}, we obtain d$\sim$ 20 nm, in good agreement with the normal metal layer thickness estimated from the McMillan model. The early flux entry will cause enhanced surface dissipation (Q slope) and a premature quench compared to cavities with homogeneous superconducting properties, consistent with the RF tests of Nb cavities with and without LTB treatements\cite{Ciovati2}. Assuming that hydrides are responsible for local weakened superconductivity the reasons for inhomogeneous precipitation on the Nb cavity inner surface remains elusive, prompting further experimental investigations.

Whereas the average gaps and distributions values measured on the LTB and N or Ti doped samples are very similar, the mean $\Gamma/\Delta$ value for the LTB sample 8-12$\pm$5$\%$ is about twice the one measured on the N doped sample 4.8$\pm$1.6$\%$. The corresponding RF tests shows that the LTB cavities does not display the pronounced anti-Q slope observed on N or Ti doped cavities. The PCTS data analysis therefore confirms that both homogeneous gap and low $\Gamma/\Delta$ values are necessary conditions for the anti-Q slope to appear. It also indicates that inelastic scattering processes are responsible for lower Q and higher dissipation occuring over the measured E$_{acc}$ range. The tunneling spectroscopy results and analysis of the superconducting properties presented in this paper can be summarized as follow:

- Homogeneous superconducting $\Delta$ values (1.5-1.6 meV) close to bulk Nb and low $\Gamma/\Delta$ values (4-5$\%$) correlate with the anti-Q slope effect.  

- Homogeneous superconducting $\Delta$ values (1.5-1.6 meV) close to bulk Nb and larger $\Gamma/\Delta$ values (8-12$\%$) measured on LTB Nb samples without HFQS correspond to RF performance with lower Q than the doped cavities; i.e. the absence of anti-Q slope, and an increased dissipation up to high accelerating fields.

- Inhomogeneous superconducting properties with $\Delta$ values significantly smaller than bulk Nb and large $\Gamma/\Delta$ values (8$\%$) are measured in a hot spot sample responsible for the HFQS. 

The occurence of tunneling spectra with kondo peaks around the Fermi level suggest localized magnetic moments as a microsocpic contributor to the inelastic scattering parameter $\Gamma$. It is striking to notice that magnetic impurities have also been found at the surface of superconducting planar resonators operating at similar frequencies (2-10 GHz) and are thought to be a signature of two level systems (TLS)\cite{TLS-1, TLS-2} responsible for the 1/f magnetic flux noise and the devices performance limitations. Several models have been developed to explain the performance limitation of these resonators based on a coupling between the localized spins present within the dielectrics layers\cite{TLS-2} or adbsorbed on the surface, the superconductor and the associated Kondo effect\cite{TLS-3}. It is therefore possible that similar dissipative mechanisms mediated by TLS could be affecting SRF cavity performances; recent RF measurements on 1.3 GHz Nb cavities revealed a decrease followed by a saturation of Q at very low accelerating fields\cite{Romanenko2} that is consistent with TLS-induced losses originating from the native niobium oxides.

\section{Conclusion}

In conclusion, we have shown by means of tunneling spectroscopy that the HT doping treatments provides an efficient technological pathway towards homogeneous, near ideal superconducting properties on the niobium surface. Dilute concentration of interstitials such as N or Ti has several beneficial effects, including the trapping of hydrogen and the formation of a thicker, less defective oxide layer.  Thus the deleterious effects of non-superconducting phases as well as magnetic moments and other potential TLS are significantly reduced.  This allows the observation of potential non-equilibrium phenomena of superconductors in strong RF fields (e.g. Gurevich\cite{Gurevich}) and the resulting anti-Q slope, to be observed. In contrast, the presence of magnetic impurities and weakened superconductivity spots originating from normal metal clusters near the surface, correlate with the strong temperature increase and dissipation occuring at high accelerating fields in BCP and EP Nb samples.  The tunneling spectroscopy results presented in this work further exemplify the need for TS as a predictive tool for future surface treatments aimed at improving superconducting resonnator performances.

\section{Acknowlegment}
This work was supported by the department of energy, Office of Sciences, Office of High Energy Physics, early career award FWP 50335 and the use of the APS by the U.S. Department of Energy, Office of Science, operated under Contract DE-AC02-06CH11357 by UChicago Argonne, LLC.

\bibliography{NDoping}

\end{document}